\documentclass[letterpaper,twocolumn,aps,prb,superscriptaddress,showpacs]{revtex4}

\usepackage{graphicx}
\begin{document}
\title{Vibrational Properties of Nanoscale Materials: From
Nanoparticles to Nanocrystalline Materials}

\author{R. Meyer}\email{ralf@thp.uni-duisburg.de}
\affiliation{D\'{e}partement de physique et Groupe de Recherche en
Physique et Technologie des Couches Mince (GCM), Universit\'{e} de
Montr\'{e}al, C.P. 6128 Succursale Centre-ville, Montr\'{e}al
(Qu\'{e}bec) H3C 3J7, Canada}
\affiliation{Institut f\"{u}r Physik,
Universit\"{a}t Duisburg-Essen, Lotharstra{\ss}e 1, D-47048
Duisburg, Germany}
\author{Laurent J. Lewis}
\affiliation{D\'{e}partement de physique et Groupe de Recherche en
Physique et Technologie des Couches Mince (GCM), Universit\'{e} de
Montr\'{e}al, C.P. 6128 Succursale Centre-ville, Montr\'{e}al
(Qu\'{e}bec) H3C 3J7, Canada}
\author{S. Prakash}
\affiliation{Jiwaji University, Gwalior, 474\,011, India}
\author{P. Entel}
\affiliation{Institut f\"{u}r Physik,
Universit\"{a}t Duisburg-Essen, Lotharstra{\ss}e 1, D-47048
Duisburg, Germany}

\begin{abstract}
The vibrational density of states (VDOS) of nanoclusters and
nanocrystalline materials are derived from molecular-dynamics
simulations using empirical tight-binding potentials. The results show
that the VDOS inside nanoclusters can be understood as that of the
corresponding bulk system compressed by the capillary pressure. At the
surface of the nanoparticles the VDOS exhibits a strong enhancement at
low energies and shows structures similar to that found near flat
crystalline surfaces. For the nanocrystalline materials an increased
VDOS is found at high and low phonon energies, in agreement with
experimental findings. The individual VDOS contributions from the
grain centers, grain boundaries, and internal surfaces show that, in the
nanocrystalline materials, the VDOS enhancements are mainly caused
by the grain-boundary contributions and that surface atoms play only a
minor role. Although capillary pressures are also present inside the
grains of nanocrystalline materials, their effect on the VDOS is
different than in the cluster case which is probably due to the
inter-grain coupling of the modes via the grain-boundaries.
\end{abstract}
\date{June 2, 2003} 
\maketitle

%
%
\section{Introduction}
The current interest in nanotechnology makes nanoclusters as well as
nanocrystalline materials the subject of many research activities.
Nanoclusters, i.e., particles with diameters in the nanometer range
containing between a few hundreds and hundreds of thousands of atoms,
are the building blocks of so-called cluster-assembled materials. An
understanding of the clusters is therefore necessary in order to
control manufacturing and properties of the resulting materials. 
Nanocrystalline materials are polycrystalline materials with grain
sizes below 100\,nm. The interest in these materials is motivated by
the fact that the small grain sizes give nanocrystalline materials
unique physical properties which are different from those of their
coarser-grained counterparts. A general introduction to both,
nanoclusters and nanocrystalline materials can be found in
Ref.~\onlinecite{Edelstein:96a}.

One possible method for the synthesis of nanocrystalline materials is
the inert gas condensation technique. In this process, nanoparticles
produced from the gas phase are deposited on a substrate and
subsequently \textit{in-situ} compacted and 
sintered.\cite{Edelstein:96a, Siegel:88a} This technique as well as
the similarity of the dimensions of nanoclusters and the grains in
nanocrystalline materials provide a kind of link between both systems
and several authors have compared vibrational properties 
calculated theoretically for clusters with 
experimental results of nanocrystalline
materials.\cite{Kara:98a,Hu:01a} In this work, we present results for
the vibrational density of states (VDOS) in nanoclusters and
nanocrystalline materials obtained from molecular-dynamics simulations
using the empirical tight-binding potentials of Cleri and
Rosato.\cite{Cleri:93a} The fact, that we have derived the VDOS for
both types of systems with the same model potential, enables us to
demonstrate that it is impossible to describe the VDOS of
nanocrystalline materials in terms of the VDOS of the clusters since
the dominating effect are different in the two kinds of systems. While
the VDOS of nanoparticles is governed by the large number of surface
atoms and the presence of a rather homogeneous compressive capillary
pressure, the VDOS of nanocrystalline materials is dominated by large
volume proportion of grain-boundary atoms and the coupling of the
individual grains via the grain boundaries.

A crucial point for the simulation of nanocrystalline materials is the
elaboration of an appropriate model with a realistic grain-boundary
structure. The most widely used approaches to generate models of
nanocrystalline materials are the controlled solidification of a
liquid used by Phillpot, Wolf and
Gleiter\cite{Phillpot:95b,Phillpot:95c} and the Voronoi construction
employed by Chen,\cite{Chen:95a} Schi{\o}tz, Di Tolla and
Jacobsen,\cite{Schiotz:98a,Schiotz:99a} as well as Van~Swygenhoven and
Caro.\cite{VanSwygenhoven:97a,VanSwygenhoven:98a} However, both of
these methods give fully dense systems with negligible amounts of free
volume, whereas the experimentally observed densities vary between 75
and 97\,\% of the density of their coarser grained
counterparts.\cite{Nieman:91a,Siegel:96a} For this reason we have
employed another approach in the preparation of our nanocrystalline
model systems. The models used for this work were obtained from
simulations of the consolidation of nanoparticles into nanocrystalline
materials. This approach mimics the experimental preparation of
nanocrystalline materials by inert gas condensation and leads to
porous systems with realistic densities. The reason why this method is
not widely used in the literature is that the simulation of the
consolidation is computationally very demanding. Early applications of
this technique\cite{Liu:94a,Zhu:95a} were therefore strongly limited
in the size of the simulated systems.  However, the rapidly growing
capabilities of modern parallel computers made it possible for us to
simulate the consolidation of model systems with more than one million
atoms.

Experimentally, it has been found that nanocrystalline metals exhibit
an anomalous increase of the VDOS at low energies.\cite{Trampenau:95a,
Fultz:96a,Frase:98a,Bonetti:00a,Stuhr:98a} Similar enhancements have
been observed in computer simulations.\cite{Wolf:95a, Derlet:01a} In
Ref.~\onlinecite{Derlet:01a}, a nickel model based on the Voronoi
construction has been compared with an early version of our model for
nanocrystalline copper. From this comparison of a fully dense model
with a porous model, the dominating effect of the grain-boundary atoms
on the VDOS of nanocrystalline metals was deduced. We obtain the same
results from a comparison of the VDOS of one model system derived at
different stages of the densification.

For a discussion of the vibrational properties of nanoparticles it is
useful to distinguish between the surface and the core of the
particles. In their work on the vibrational properties of clusters,
Kara and Rahman reported that the VDOS of the inner atoms of a cluster
is very similar to the VDOS of the corresponding bulk material but
shifted to higher energies.\cite{Kara:98a} Later, it has been shown
that this shift is caused by the capillary pressure acting inside
nanoparticles.\cite{Meyer:02a} Naturally, the distinction of surface
and core contributions requires a suitable definition of these
regions.  In Ref.~\onlinecite{Kara:98a}, the core (or inner part) of
the cluster was defined to consist of all atoms with the bulk
coordination number $Z=12$. The problem with this definition is that a
lot of the atoms with $Z=12$ near the surface show a very
different behavior than those deep inside the cluster. For this reason
it seems more reasonable to use a definition that gives smaller and
more homogeneous particle cores. This was done in
Ref.~\onlinecite{Meyer:02a} by considering the second invariant $J_2$ of
the local stress tensor, which turns out to be fairly constant inside
the particles. The disadvantage of this method, however, is that it
contains some arbitrariness since the onset of the deviation of $J_2$
has to be determined by manual inspection. Here, we propose another
criterion based on the common neighbor analysis
(CNA)\cite{Honeycutt:87a} which yields very similar results than the
method used in Ref.~\onlinecite{Meyer:02a} without arbitrariness.

For a macroscopic spherical particle of radius $R$ it is known that
inside the sphere acts a capillary pressure $p$ of magnitude
\begin{equation}
p=\frac{2 \tau}{R},
\label{EqKelvin}
\end{equation}
where $\tau$ is the (average) surface stress of the material.
Previously, it has been shown for two different types of embedded-atom
method potentials that this equation remains valid for particles in
the nanometer regime.\cite{Swaminarayan:94a,Meyer:02a} In the present
work we find that, for the empirical tight-binding potentials of Cleri
and Rosato,\cite{Cleri:93a}
 Eq.~(\ref{EqKelvin}) holds for particles
with diameters above a critical size of approximately 2.5\,nm.
Moreover, we report the values of the surface stress $\tau$ for Ag,
Au, Cu, Ni, and Pt obtained from our simulations with the help of
Eq.~(\ref{EqKelvin}).

%
%
\section{Details of the Calculations}
\subsection{General}
In this article we report on results of molecular-dynamics simulations
employing the empirical tight-binding second moment potentials of
Cleri and Rosato\cite{Cleri:93a} for the calculation of the
interatomic forces. These potentials, which are fitted to the
experimental values of cohesive energy, lattice parameters and elastic
constants, have been successfully applied in many works concerning
clusters\cite{Lopez:95a,Celino:96a,Palacios:99a,Sun:01a,Darby:02a,
Michaelian:02a,Rexer:02a,AguileraGranja:02a} and nanocrystalline
materials.\cite{Celino:95a,Celino:95b,VanSwygenhoven:97a,Derlet:01a,
Derlet:02a,Samaras:02a} The equations of motion in our simulations
were integrated with the help of the velocity form of the Verlet
algorithm\cite{Allen:91a} using a time-step $h = 2\,\mbox{fs}$. All
simulations were carried out at a room temperature
($T=300\,\mbox{K}$). For systems containing more than a few thousands
of atoms it becomes unfeasible to calculate the frequencies of the
vibrational modes by diagonalization of the dynamical matrix. For this
reason we derived phonon density of states $\rho(\hbar \omega)$ from
the Fourier transform of the velocity autocorrelation function
$\langle \mathbf{v}(t)\mathbf{v}(0) \rangle$ obtained from the
simulations using the methods described in the book of Allen and
Tildesley.\cite{Allen:91a} The knowledge of the VDOS makes it possible
to calculate a number of thermodynamic properties within the harmonic
approximations. Here, we use the VDOS to derive the specific heat per
atom $c$ of the simulated systems with the help of
\begin{equation}
  c = 3\,k_{\mathrm{B}} 
      \int_0^\infty 
      \left ( \frac{\hbar \omega}{k_{\mathrm{B}}T} \right )^2
      \rho(\hbar \omega) \,
      \frac{\mathrm{e}^{\hbar \omega / k_{\mathrm{B}}T}}
      {(\mathrm{e}^{\hbar \omega / k_{\mathrm{B}}T}-1)^2} \;
      \mathrm{d}\omega .
\label{EqCp}
\end{equation}

\subsection{Nanoparticles}
For the investigation of nanoparticles we have simulated Ag, Au, Cu,
Ni, and Pt nanoparticles with diameters in the range of 2 to
10\,nm. As in our previous work,\cite{Meyer:02a} the simulations were
started with spherical configurations of 531, 791, 1205, 1865, 3043,
8247, 17957, and 36417 atoms on an fcc lattice. During the
initialization, care was taken to achieve vanishing total momenta and
angular momenta of the systems. These configurations were then
equilibrated at $T=300\,$K over a period of 50\,000 simulation steps
before the main simulation runs of another 50\,000 steps.

After the simulations we used the CNA to group the atoms in the final
configurations into four classes: perfect fcc (PFCC) atoms with an fcc
environment up to the fourth neighbor shell, good fcc (GFCC) atoms
with an fcc environment in the first neighbor shell, surface atoms
(SURF) with coordination number $Z<10$, and other atoms (OTHR). It
turns out that for the 531 and 791 atom clusters the PFCC group is
identical to the core found with the previously-used
method.\cite{Meyer:02a} Motivated by this finding we identify in what
follows the particle cores with the PFCC atoms. The advantages of this
definition are that, on the one hand, it leads to much more homogeneous
cores than the simple coordination number criterion of
Ref.~\onlinecite{Kara:98a} while, on the other hand, it avoids the
arbitrariness of the method employed in Ref.~\onlinecite{Meyer:02a}.

The core definition explained in the preceding paragraph enables us 
to calculate the capillary pressure $p$ acting inside the particles as
a sum over the local pressures at the core atoms: 
\begin{equation}
  p = \frac{1}{3 N_\mathrm{c} \Omega_\mathrm{c}} {\sum_{i}}
      \sum_{\alpha} \left \langle m_i \left ( v_i^\alpha \right )^2
      +  {\sum_{j\ne i}} F_{ij}^{\alpha} r_{ij}^{\alpha} \right 
      \rangle.
\label{EqCap}
\end{equation}
Here, $m_i$ is the mass of particle $i$ while $v_i^\alpha$,
$F_{ij}^\alpha$, and $r_{ij}^\alpha$ are the Cartesian components
($\alpha \in x,y,z$) of the velocity of particle $i$, the force acting
between particles $i$ and $j$ and their distance, respectively.
Furthermore, angular brackets denote thermal averaging and the primed
sums run over all core particles, i.e., all PFCC particles.
Finally, $N_\mathrm{c}$ is the number of particles in the core and
$\Omega_\mathrm{c}$ the average atomic volume of the core particles
which we have derived from the mean nearest neighbor distance
$\bar{r}_\mathrm{nn}$ inside the core as $\Omega_\mathrm{c} =
\frac{1}{2}\sqrt{2} \bar{r}_\mathrm{nn}^3$.

In order to compare the capillary pressures obtained from
Eq.~(\ref{EqCap}) with the prediction of Eq.~(\ref{EqKelvin}) the
radius $R$ of the particle has to be known. Since, to our knowledge,
there is no exact and unique definition of the radius of a cluster, we
adopted here the previously used simple scheme for this
purpose.\cite{Meyer:02a} Thus, the cluster radius $R$ is
defined to be equal to the largest distance of a particle with
coordination number $Z \ge 8$ from the particles center of mass.

Finally, we have derived the partial contributions of the four groups
of atoms (PFCC, GFCC, OTHR, and SURF) to the total VDOS of the
$\mathrm{Cu}_{791}$ particle. This was done by averaging the
individual VDOS contributions obtained from the
velocity autocorrelation function of 1000 simulation runs over a
period of 25\,000 steps, each. Such a long simulation time was
necessary to compensate for the poor statistics caused by the small
number of particles in the system.

%
%
\subsection{Nanocrystalline Materials} 
\label{SecDetailsNC}
As stated in the introduction the nanocrystalline model configurations
used in this work were obtained from simulations of pressure-assisted
nanoparticle sintering. With this method we have constructed models
for nanocrystalline Cu, Ni, and Ag. 

The starting point of the sintering simulations were configurations
composed of the equilibrated configurations of clusters described in
the preceding section. Randomly oriented copies of these clusters were
placed in a cubic box with twice the volume of a bulk system with the
same number of atoms. In order to avoid overlapping of the clusters in
the initial configurations, the positions of the clusters were
determined with the help of a Monte-Carlo procedure. In this
procedure, the clusters were regarded as point particles interacting
by a purely repulsive pair potential.\cite{Weeks:71a} For each pair
of particles the range of the potential was chosen to be larger than
the sum of the particle radii plus the interaction range of the Cleri
and Rosato potential used in the molecular-dynamics simulations.  Due
to this construction the interaction energy in the Monte-Carlo
simulation is exactly zero if all clusters are well separated and
positive otherwise. Starting from arbitrary positions the clusters were
then moved in trial steps which were only accepted if they resulted in
a lower total interaction energy. This procedure was continued until
the total energy remained zero for 100 continuous trial moves of all
clusters. Details of the numbers and sizes of the clusters used for
the generation of the initial configuration are given in
Table~\ref{TabBuild}.
\begin{table}
\caption{Composition of the starting configurations for the
nanoparticle sintering simulations. Each line specifies the number $n$
of clusters containing $N$ atoms and their diameters $d$ in nm for Cu,
Ni, and Ag. The total number of atoms is 1\,007\,120 in all three
cases.} 
\label{TabBuild}
\begin{ruledtabular}
\begin{tabular}{rrrrr}
\multicolumn{1}{c}{$n$} & \multicolumn{1}{c}{$N$} & 
\multicolumn{1}{c}{$d_\mathrm{Cu}$} & \multicolumn{1}{c}{$d_\mathrm{Ni}$} &
\multicolumn{1}{c}{$d_\mathrm{Ag}$} \\
\noalign{\smallskip}\hline\noalign{\smallskip}
20 &  3043 & 4.0 & 3.9 &  4.5 \\
60 &  8247 & 5.6 & 5.5 &  6.3 \\
15 & 17957 & 7.3 & 7.2 &  8.3 \\
 5 & 36417 & 9.3 & 9.1 & 10.5 \\
\end{tabular}
\end{ruledtabular}
\end{table}

After their generation the starting configurations were simulated at
constant temperature and pressure using the Parinello-Rahman
scheme\cite{Parrinello:80a} restricted to orthogonal simulation boxes
and the Nos\'{e}-Hoover thermostat method.\cite{Hoover:85a} The
application of the latter method was necessary to remove the excess
energy released during the densification from the system. The
simulations were started with a simulation run at zero pressure over
50\,000 simulation steps (100\,ps). Another 50\,000 simulation steps
were then performed at a hydrostatic pressure of 100\,MPa.  This
predensification phase was followed by three consecutive simulation
runs at pressures of 0.7, 1.4, and 2.1\,GPa with a length of 100\,000
steps (200\,ps), each. In the last step, the intermediate
configurations resulting from the simulations at 0.7, 1.4, and
2.1\,GPa were slowly relaxed to zero pressure, again. This was done in
a simulation where the pressure was reduced by 175\,MPa after each
2500 simulation steps. Eventually, the simulations of the relaxed
systems were continued for 10\,000 steps at zero pressure in order to
diminish possible perturbations caused by the stepwise reduction of
the pressure. The nominal grain sizes of the final configurations,
calculated under the optimistic assumption that each of the initial
particles forms one grain in the resulting material, are 7\,nm in the
case of Ag and 6\,nm for Cu and Ni. The actual values, however, will
be somewhat higher since during the compaction surely some of the
particles joined to form single grains.

The models resulting from the relaxation of the the systems after the
sintering under different pressures enabled us to analyze the VDOS of
the nanocrystalline metals at different densities. We did this
analysis for the models of nanocrystalline copper obtained from
maximum pressures of 0.7 and 2.1\,GPa. For these configurations we
derived the VDOS from the velocity autocorrelation function averaged
over 10 consecutive simulation runs with a length of 10\,000 steps,
each. In addition to the total VDOS, we calculated the partial
contributions of the following groups of atoms, identified by a CNA
analysis of the averaged atomic positions: atoms with a perfect fcc or
hcp environment up to at least the fourth neighbor shell (PFCC, PHCP),
good fcc or hcp atoms with an fcc or hcp like environment in the first
and more neighbor shells (GFCC, GHCP), surface atoms with coordination
number $Z<10$ (SURF) and grain boundary atoms (GB) not belonging to
any of the other groups.

Nanocrystalline materials inherently contain a certain amount of
excess volume in the form of open or closed pores. In order to analyze
the free volume in our model configurations we followed a procedure
similar to that described by Campbell et al.:\cite{Campbell:99a} we
subdivided the simulation boxes into a regular grid of voxels (the 3-d
equivalent of a pixel) and performed an analysis of the clusters of
empty voxels. However, in contrast to Campbell et al., we did not use
simple cubic voxels for this analysis but the Wigner-Seitz cell of the
fcc lattice since this choice gives considerably less anisotropic
voxels. The edge length of the voxels were chosen to be 5.75, 5.60,
and 6.50\,a.\,u.\ for Cu, Ni, and Ag, respectively. Incomplete voxels,
resulting from the incommensurability of the simulation boxes with
respect to the voxel lattice, were excluded from this analysis.

%
%
\section{Results}
\subsection{Nanoparticles}
\label{SecResultsCl}
\begin{figure}[t]
\centerline{\includegraphics[width=7.5cm]{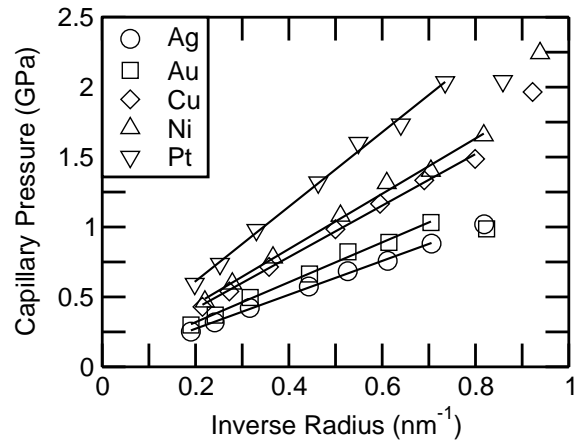}}
\caption{Capillary pressure inside fcc metal nanoclusters as a function of the
inverse cluster radius $R^{-1}$. The lines are the results of linear
fits to the data of all but the smallest particles.}
\label{FigPress}
\end{figure}
In Fig.~\ref{FigPress} we present the capillary pressures inside the
nanoparticle configurations calculated from our simulations with the
help of Eq.~(\ref{EqCap}) and fitted to the simple model of
Eq.~(\ref{EqKelvin}). From this figure it is clear that
Eq.~(\ref{EqKelvin}) accounts for the capillary pressure in all
configurations except the smallest ones. Thus, within the Cleri and
Rosato potentials the scaling regime where the capillary pressure
scales monotonously with the system size extends down to particle
sizes of 2 to 3\,nm. For the smallest clusters below this size,
Fig.~\ref{FigPress} shows strong deviations from the linear behavior
of Eq.~(\ref{EqKelvin}) whose signs and magnitudes vary strongly
between the elements.

\begin{table}[b]
\caption{Calculated and experimental values of the surface stress
$\tau$ in various fcc metals.}
\label{TabTau}
\begin{ruledtabular}
\begin{tabular}{llcllc}
 & Ag & Au & Cu & Ni & Pt \\
\noalign{\smallskip}\hline\noalign{\smallskip}
$\tau_\mathrm{sim}(\mathrm{N/m})$ & 1.2 & 1.4 & 1.8 & 2.0 & 2.7 \\
$\tau_\mathrm{exp}(\mathrm{N/m})$ & 1.4\footnotemark[1] &
        1.2\footnotemark[2]--3.2\footnotemark[3] & 0.0\footnotemark[4] & 
        & 2.6\footnotemark[4] - 4.4\footnotemark[3] \\
\end{tabular}
\end{ruledtabular}
\footnotetext[1]{Ref.~\onlinecite{Wasserman:70a}}
\footnotetext[2]{Ref.~\onlinecite{Mays:68b}}
\footnotetext[3]{Ref.~\onlinecite{Solliard:85a}}
\footnotetext[4]{Ref.~\onlinecite{Wasserman:72a}}
\end{table}
The fact that the capillary pressure inside the simulated
nanoparticles follows the linear relationship of Eq.~(\ref{EqKelvin})
makes it possible to derive the average surface stress $\tau$ for the
different elements. The results are presented in Table~\ref{TabTau}
together with some experimental data. Two points have to be noted
here. First, in the case of Cu, there is a remarkable difference
between our result and the experimental value. However, compared to
the other metals, it is the experimental value which seems to be
exceptionally low. In fact, in Ref.~\onlinecite{Wasserman:72a}, two
different surface stresses of 0.0 and 5.3\,N/m have been
reported. These values follow from lattice constant measurements using
two different electron diffraction spots. The authors of
Ref.~\onlinecite{Wasserman:72a} reject the higher value since the
corresponding observations of the (111) peak were less reliable than
those of the (220) peak which gave the lower value. The second point
to be noted is the strong scatter in the experimental data of Au and
Pt. One possible reason for this are the relatively large error bounds
of up to 1\,N/m. In addition to this, Solliard and Flueli
report that their values for the surface stress of Au have been
increased by the occurrence of structural changes of the
clusters.\cite{Solliard:85a} However, consistent calculation of the
surface stress from Eq.~(\ref{EqKelvin}) is only possible if all data
points correspond to the same structure. Taking all these experimental
uncertainties into account, the surface stresses resulting from our
simulations are in good agreement with the experimental data. In order
to shed more light on the case of Cu, independent measurements or
ab-initio calculations would be highly desirable.

\begin{figure}
\centerline{\includegraphics[width=7.5cm]{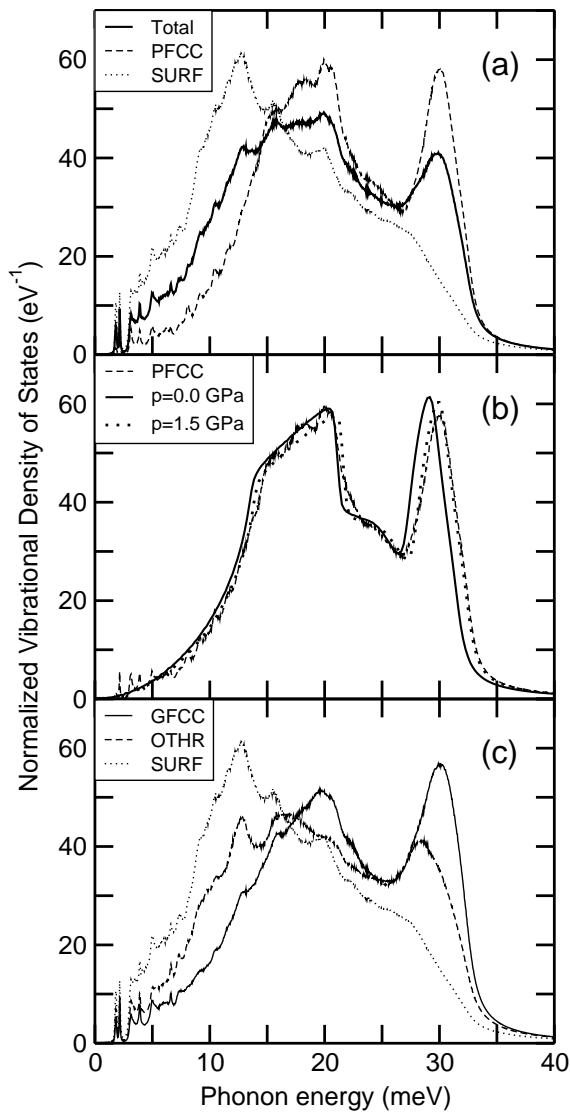}}
\caption{Normalized VDOS of the $\mathrm{Cu}_{791}$ cluster. (a) Total
VDOS compared to the surface (SURF) and core (PFCC) contributions; (b) Core
part of the VDOS compared to the VDOS of crystalline fcc Cu under
pressures of 0. and 1.5\,GPa; (c) Comparison of the three non-core VDOS
contributions.}
\label{FigCluster}
\end{figure}
The VDOS of the $\mathrm{Cu}_{791}$ particle and its partial
contributions are shown in Fig.~\ref{FigCluster}.  In the upper panel,
Fig.~\ref{FigCluster}(a), the contributions from surface and core
atoms are compared to the total VDOS, demonstrating the huge
differences between both contributions. While the core contribution
has two clearly separated peaks corresponding to the longitudinal and
transversal modes of the fcc VDOS, the surface contribution has only
one peak at energies below the transversal peak of the core and a
shoulder at the position of the longitudinal peak. The form of the
surface VDOS resembles closely the VDOS found near flat
surfaces.\cite{Black:90a} From these results, it is clear that the
surface atoms lead to a significant increase of the VDOS of
nanoparticles at low energies whereas they reduce the VDOS at the
upper end of the spectrum.

In Fig.~\ref{FigCluster}(b) the particle's core VDOS is given together
with the VDOS of crystalline fcc Cu under pressures of 0. and
1.5\,GPa, where 1.5\,GPa corresponds to the capillary pressure acting
on the cluster's core. This comparison reveals, in accordance with
previous findings,\cite{Meyer:02a} that the VDOS inside the core of a
nanoparticle is well accounted for by the bulk system under
pressure. In particular, the capillary pressure gives rise to a small
shift of the longitudinal peak to higher energies, thereby slightly
increasing the high-energy VDOS.

Finally, Fig.~\ref{FigCluster}(c) shows the three non-core
contributions to the VDOS of the $\mathrm{Cu}_{791}$ particle. From
this figure, it can be seen that there is a gradual transition from
the GFCC to the SURF contribution which reflects the geometry of the
three groups of atoms: The perfectly fcc ordered core is directly
surrounded by a shell of GFCC atoms. The GFCC atoms in turn are
surrounded by subsurface atoms with coordination numbers $Z=10,11$
followed by the outermost layer of surface atoms with $Z<10$.

\subsection{Nanocrystalline Materials}
During the simulations of nanoparticle sintering, the systems evolved
from a gas of individual particles via a loose agglomerate of
particles to a dense solid material. An important parameter of this
compaction process is the relative density defined as the actual
density of the system divided by the density of the corresponding
crystalline bulk material. From the relative density it can be seen
how much excess volume remains in the compacted system. In
Table~\ref{TabDensity} we report the relative densities of our
simulated nanocrystalline materials after the sintering pressure had
been removed. Not surprisingly, the relative density increases for all
three elements as the sintering pressure (and time) increases.
However, there are some subtle differences between the metals which
are worth noting. After compaction without application of an external
pressure Cu and Ni both show the same relative density of 77\,\%
whereas the Ag system has a considerable lower density of 70\,\%. In
contrast to this, the picture has changed at the end of our
simulations, where the Ag system has the highest density, followed by
Cu and Ni.
\begin{table}[b]
\caption{Relative densities of the nanocrystalline configurations
after removal of the maximum sintering pressure $p_\mathrm{max}$ (GPa).}
\label{TabDensity}
\begin{ruledtabular}
\begin{tabular}{cccc}
$p_\mathrm{max}$ & Cu & Ni & Ag \\
\noalign{\smallskip}\hline\noalign{\smallskip}
0.0 & 0.77 & 0.77 & 0.70 \\
0.7 & 0.89 & 0.85 & 0.90 \\
1.4 & 0.94 & 0.89 & 0.96 \\
2.1 & 0.96 & 0.92 & 0.98 \\
\end{tabular}
\end{ruledtabular}
\end{table}

The different behavior of the three metals during the compaction
described in the previous paragraph can be understood from the
differences of the stacking-fault and surface energies of the three
metals. At the beginning of the compaction, when the system is not a
solid system but a loose powder, the compaction rate is mainly
determined by the surface energy of the metals which is the driving
force behind the whole sintering process. Since the surface energy is
highest for nickel (see e.\,g. Ref.~\onlinecite{Meyer:02a}) and lowest for
Ag, the densification starts fastest for Ni and slowest for
silver. However, as the compaction proceeds, further densification
requires the plastic deformation of the particles which is controlled
by the stacking-fault energy. The densities at the end of the
simulations therefore reflect the fact that (experimentally as well as
in the model) the stacking-fault energy in Ag is lower than in Cu
which in turn is much lower than that in Ni.\cite{Meyer:02b} The
influence of the stacking-fault energy on the compaction rate can
also be seen from the number of atoms with an hcp environment,
which indicate the presence of stacking faults due to plastic
deformation.  Among the configurations belonging to the first line of
Table~\ref{TabDensity} the Ag system has the lowest number of hcp
atoms. This means that the system has undergone less plastic
deformation than the other configurations and explains the rather low
relative density of the silver system at this point. At the end of the
simulations, however, the situation has reversed. Due to its high
stacking-fault energy, it is now the Ni configuration which has the
lowest number of hcp atoms as well as the lowest relative density.

\begin{figure}
\centerline{\includegraphics[width=7.5cm]{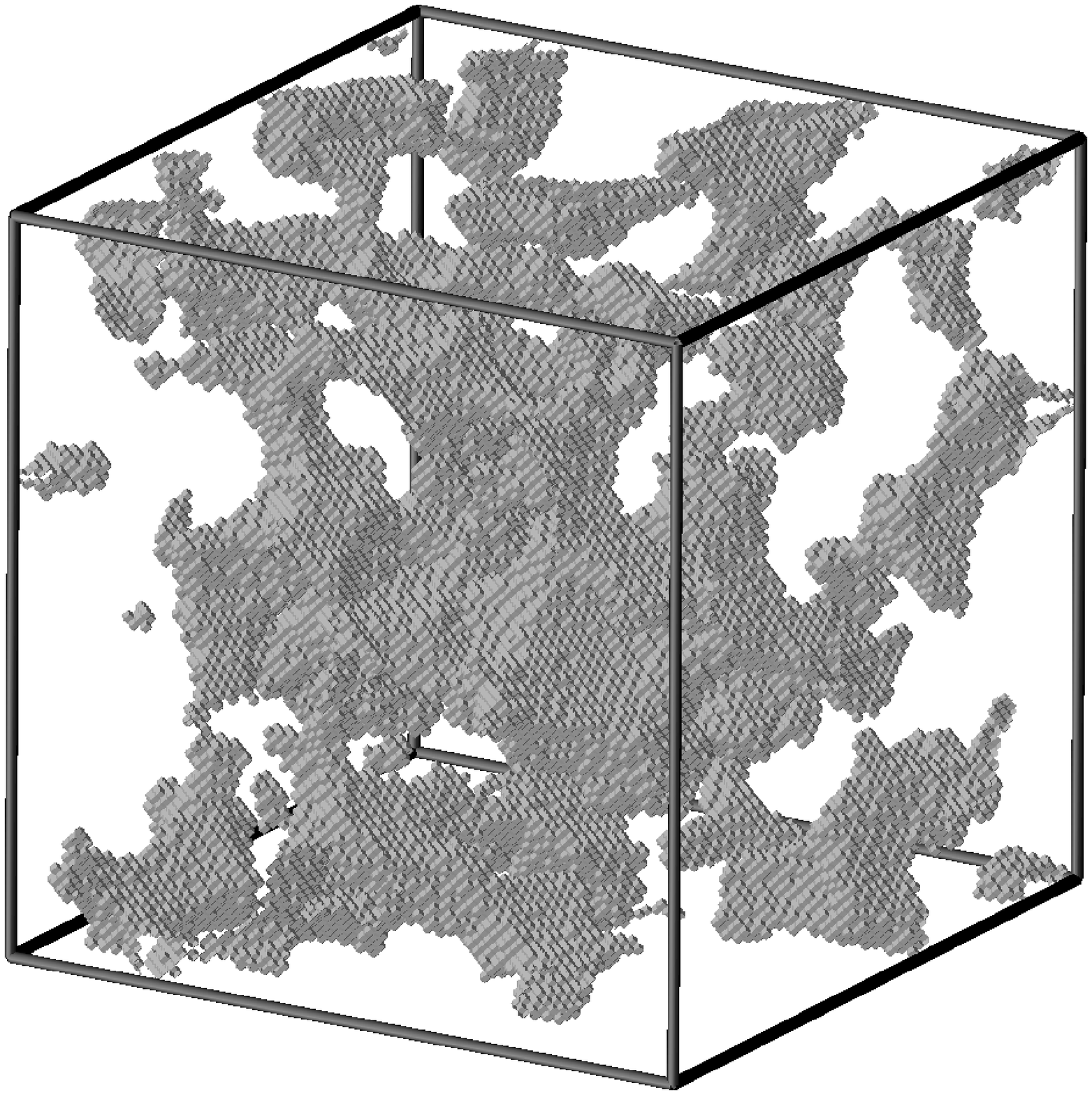}}
\centerline{\includegraphics[width=7.5cm]{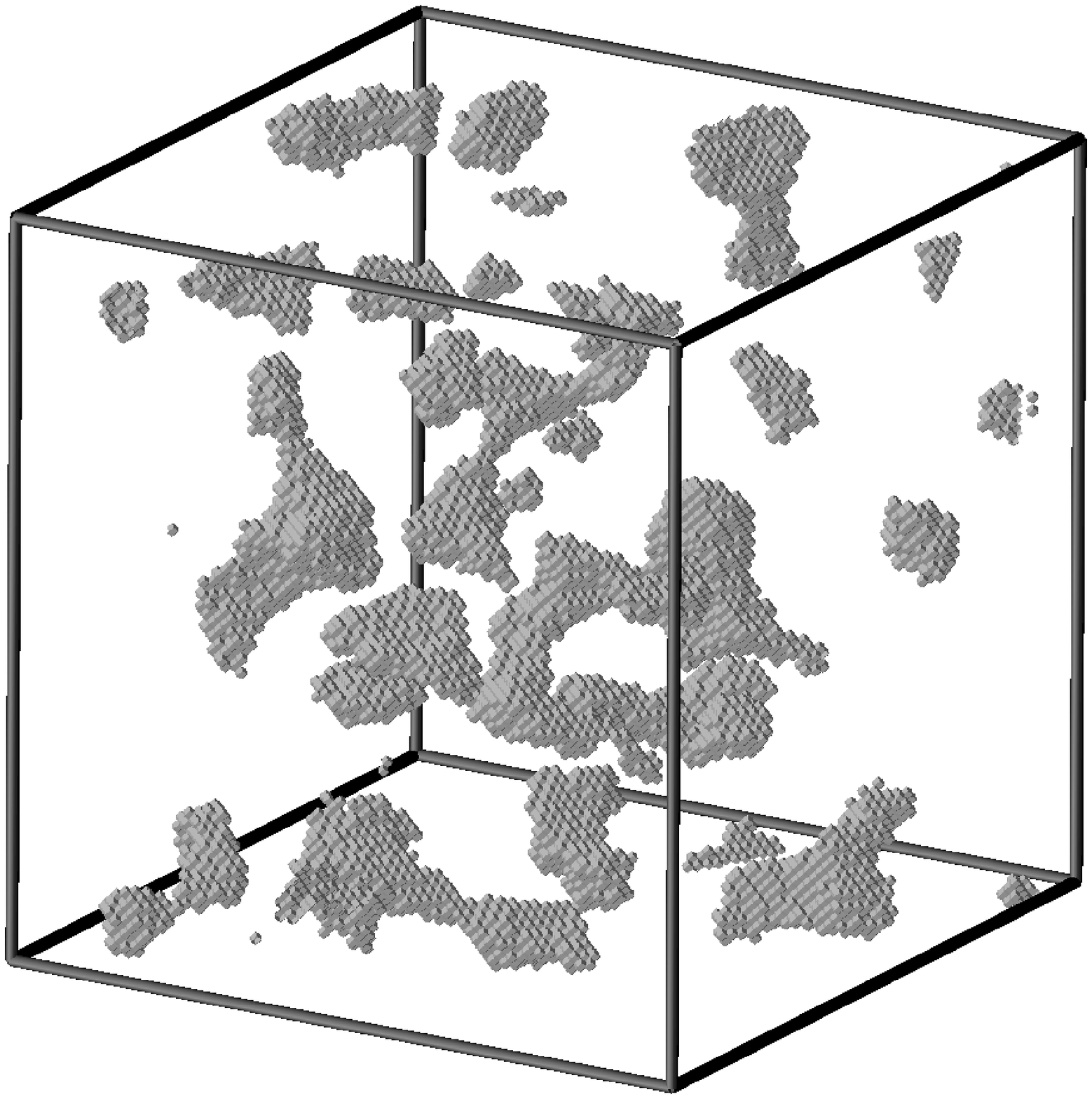}}
\caption{Free volumes in nanocrystalline Cu configurations after
sintering with relative densities of 89\,\% (top) and 96\,\%
(bottom). Clusters containing less than three voxels have been
suppressed.}
\label{FigFree}
\end{figure}
While the relative density gives information about the amount of
excess volume remaining in the systems, it tells nothing about its
structure or distribution. Such information can be gained from the
free volume analysis described at the end of
section~\ref{SecDetailsNC}. Figure~\ref{FigFree} shows the free volume
found by this method in our models of nanocrystalline copper after
sintering under pressures of 0.7 and 2.1\,GPa with relative densities
of 89 and 96\,\%, respectively. It can be seen that there is a
significant difference between the pore structures of both systems. In
the less dense system, most of the free volume is stored in one big
percolating cluster, whereas in the denser system the free volume is
distributed over several isolated clusters. This means that during the
sintering the structure of the free volume changed from an extended
network of open pores to a closed pore state. Analysis of all
configurations shows the occurrence of extended open pores in all
configurations with relative densities up to 90\,\% and closed pores
at higher densities.

\begin{figure}
\centerline{\includegraphics[width=7.5cm]{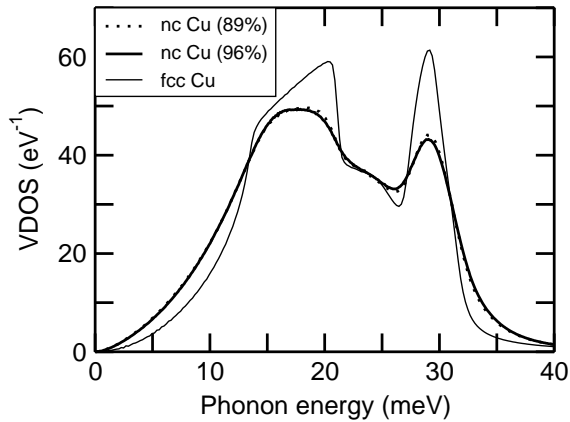}}
\caption{Normalized total VDOS of nanocrystalline (nc) copper at 89
and 96\,\% relative density together with that of crystalline (fcc)
copper.}
\label{FigTot}
\end{figure}

In Fig.~\ref{FigTot}, the total VDOS of the two Cu systems discussed
in the previous paragraph are compared with that of crystalline fcc
copper.  The VDOS of both nanocrystalline systems show strong
enhancements at high and low energies, as seen before
theoretically\cite{Derlet:01a} as well as
experimentally.\cite{Bonetti:00a} The important point here is that the
curves of both nanocrystalline systems are nearly identical, although
Fig.~\ref{FigFree} shows clearly a big difference in the amount of
internal surface area in both systems. This establishes that the
enhancements are not caused by any kind of surface effects.

\begin{figure}
\centerline{\includegraphics[width=7.5cm]{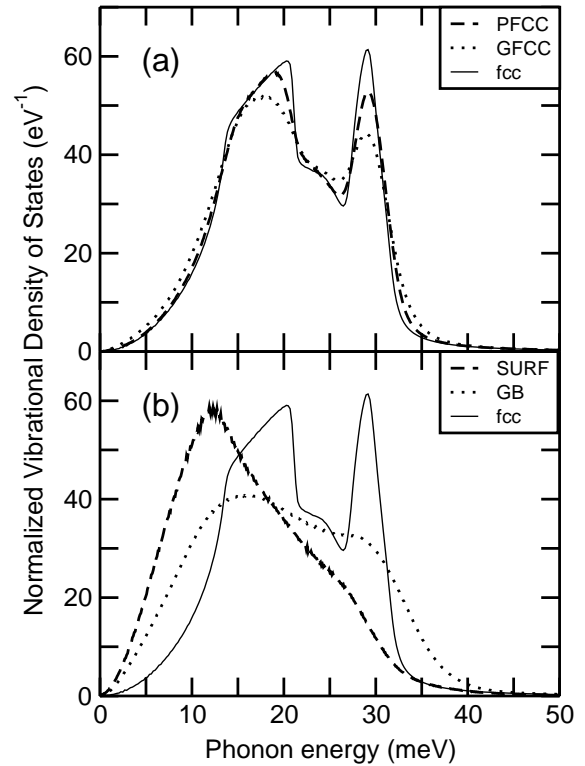}}
\caption{Normalized partial VDOS of nanocrystalline (nc)
copper at 96\,\% relative density. See text for details.}
\label{FigPart}
\end{figure}

In addition to the total density of states we also calculated the
partial contributions to the VDOS for our densest copper system. The
results of these calculations are shown in Fig.~\ref{FigPart}. In
panel (a), we compare the VDOS from atoms with perfect
(PFCC) or good (GFCC) fcc environment, i.e., atoms inside the grains,
with the crystalline VDOS. It can be seen that the VDOS inside the
grains of the nc material is very similar to that of the
crystalline bulk system.  The difference consists of a slight
broadening and a loss of the sharp structure of the peaks due the a
lack of long range order and the presence of a strong stress
field. Interestingly, the peak of the longitudinal modes is not at all
shifted with respect to the crystalline VDOS. This is striking, since
we find that a net pressure of 460\,MPa acts on the PFCC atoms forming
the core of the grains.  Although lower than the capillary pressure
acting inside the $\mathrm{Cu}_{791}$ cluster one would expect that
this pressure gives rise to a visible shift of the longitudinal
modes. A possible reason for the observed different behavior is that
the modes inside the grains are extended, i.e., they
are coupled among the grains via the grain boundaries. In that case
the looser coupling between the grains might counteract the stronger
coupling inside the grains due to the capillary pressure.

Although the broadening of the intra-grain VDOS leads to a small
increase of the VDOS at high and low energies, the effect is not
strong enough to account for the strong increase of the total VDOS
visible in Fig.~\ref{FigTot}. The biggest contribution to this
increase stems from the high amount of grain-boundary atoms which show
a very broad and nearly structureless VDOS with strong enhancements at
high and low energies (see Fig.~\ref{FigPart}(b)). In contrast to this
the VDOS of atoms at internal surfaces, also given in this figure,
shows an even stronger increase of the VDOS at low energies but no
increase in the high energy regime. This gives further support to the
observation that the contribution of the surface atoms to the observed
enhancements of the total VDOS is insignificant.

\subsection{Nanoparticles vs. Nanocrystalline Materials}
\begin{figure}
\centerline{\includegraphics[width=7.5cm]{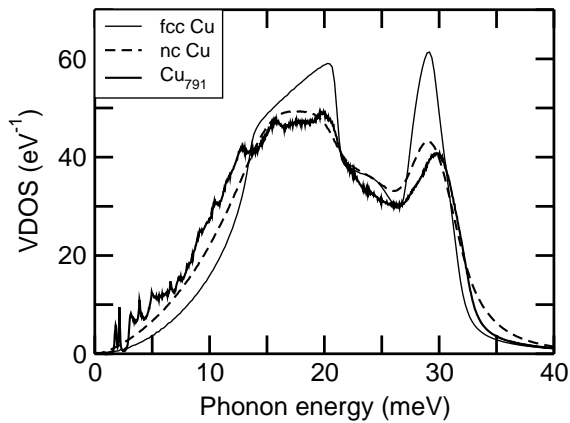}}
\caption{Normalized VDOS of a $\mathrm{Cu}_{791}$ cluster,
  nanocrystalline (nc) copper with a relative density of 96\,\% and 
  crystalline (fcc) copper.}
\label{FigDOS}
\end{figure}

Having discussed our results concerning nanoparticles and
nanocrystalline materials, we now compare directly the vibrational
properties of the two kinds of systems. For this purpose we present in
Fig.~\ref{FigDOS} the total VDOS of the $\mathrm{Cu}_{791}$ cluster
and nanocrystalline copper sintered at 2.1\,GPa together with the VDOS
of crystalline fcc copper. From this figure it can be seen that both
nanoscale systems show a broadening of the VDOS which gives rise to an
increase at high and low phonon energies. However, while for the
nanocrystalline material the increase is similar on both sides of the
spectrum, it is much stronger at low than at high energies in the case
of the cluster. The reason for this difference lies in the nature of
the atoms contributing most to the increase of the VDOS. As discussed
above, the low and high energy enhancements of the VDOS in the
nanocrystalline material are mainly due to the high number of grain
boundary atoms which show a rather symmetric broadening of the VDOS
(see Fig.~\ref{FigPart}(b)). In contrast to this it is clear from
Fig.~\ref{FigCluster}(a) that the VDOS enhancements in the cluster are
driven by two different types of atoms: The strong low energy increase
reflects the VDOS contribution of the surface atoms whereas the high
energy increase is caused by the shift of the core VDOS due to the
capillary pressure.  Finally, it can be seen from Fig.~\ref{FigDOS}
that the shift observed for the longitudinal modes in the cluster core
only leads also to a visible difference between the VDOS of the
cluster and the nanocrystalline material.

\begin{figure}
\centerline{\includegraphics[width=7.5cm]{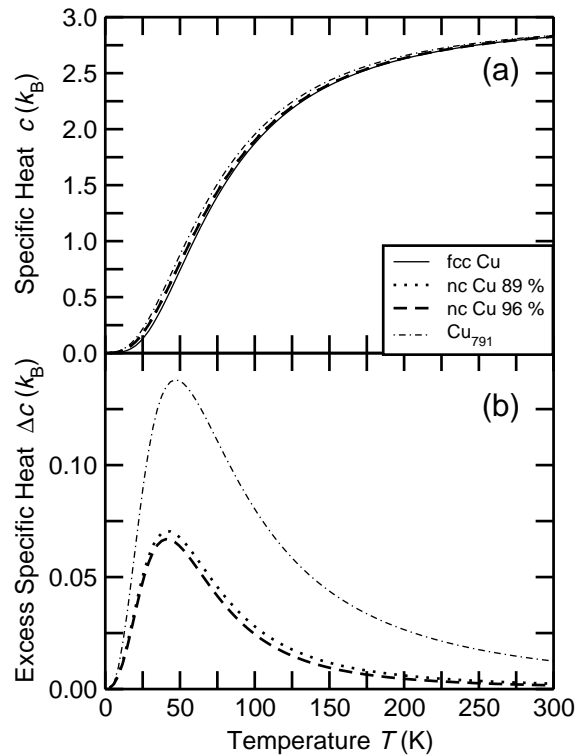}}
\caption{(a) Specific heats calculated for crystalline (fcc) copper,
  nanocrystalline copper with relative densities of 89 and 96\,\% and
  a $\mathrm{Cu}_{791}$ cluster; (b) Excess specific heat of the
  nanoscale systems in panel (a) relative to the crystalline state.}
\label{FigCp}
\end{figure}

In order to study the effect of the VDOS changes in the nanoscale
materials on other physical properties, we have calculated the specific
heat of our model systems with the help of Eq.~(\ref{EqCp}). The
results presented in Fig.~\ref{FigCp} reveal an increase of the
specific heat in the nanoscale systems at low temperatures.  All three
nanoscale systems show the maximum of the excess specific heat with
respect to the crystalline state at a temperature around 50\,K. The
height of this maximum, on the other hand, is much stronger in the
case of the nanoparticle due to the stronger increase of the VDOS at
low energies in this system. Our results here are in good qualitative
and quantitative agreement with other theoretical results for Si
nanoparticles\cite{Hu:01a} and a nanocrystalline Lennard-Jones
system.\cite{Wolf:95a} Experimentally, excess specific heats have
been observed for several nanocrystalline
metals.\cite{Rupp:87a,Tschoepe:93a,Loeffler:94a,Chen:95b,Bai:96a} At
least in some of these works,\cite{Rupp:87a,Tschoepe:93a} the measured
increase of the specific heat is considerably higher than the
theoretical results. However, it has been pointed out by Tsch\"{o}pe
and Birringer\cite{Tschoepe:93a} that large part of the measured
excess specific heat might be due to contaminations of the samples
with lighter elements. L\"{o}ffler,\cite{Loeffler:94a} on the other
hand, reports increases of the specific heat by approximately 7\,\% in
nanocrystalline Pd, which compares well to our results although he
finds the maximum of the increase at a temperature slightly above
150\,K.

%
%
\section{Summary and Discussion}
In this work we have studied nanoparticles and nanocrystalline
materials made of fcc metals with the help of molecular-dynamics
simulations based on the empirical potentials of Cleri and
Rosato.\cite{Cleri:93a}

In order to distinguish between core and surface parts of the
nanoparticles, we made use of the common neighbor
analysis\cite{Honeycutt:87a} and defined the core of the particles as
the set of atoms with a perfect fcc environment. With the help of this
criterion we studied the capillary pressure acting inside spherical
nanoparticles of Ag, Au, Cu, Ni, and Pt with diameters between 2 and
10\,nm. We find that the capillary pressure follows the same behavior
as macroscopic spherical particles down to sizes of approximately
2.5\,nm. These findings allowed us to derive the average surface
stress values for Ag, Au, Cu, Ni, and Pt predicted by the potentials
of Cleri and Rosato. Except for Cu our results agree with the
available experimental data.

Our investigations of the VDOS of a $\mathrm{Cu}_{791}$ nanoparticle
show, in agreement with our previous findings,\cite{Meyer:02a} that
the VDOS inside the core of the cluster is identical to the VDOS of
fcc copper under the capillary pressure. One particular effect of the
capillary pressure is a shift of the longitudinal modes peak to higher
energies. Outside the core the VDOS changes gradually to the typical
surface behavior of the outermost atoms. Compared to the VDOS of the
bulk material, the cluster VDOS appears to be slightly broadened with
an increased VDOS at high and low energies due to the core and surface
atoms, respectively.

The model configurations for nanocrystalline Ag, Cu, and Ni used in
this work have been obtained from extensive simulations of the
sintering of nanoparticles. These calculations resulted in
nanocrystalline configurations of the three metals with different
relative densities in the range 70 to 98\,\%. The development of
the relative density of the three metals during the compaction process
suggests that the compaction rate of the material is initially
governed by the surface energy of the material. As plastic deformation
becomes more and more important during the compaction, the
stacking-fault energy starts to dominate the process. It should be
noted, however, that the role of other mechanisms like surface or
grain-boundary diffusion might be underestimated by our simulations
since such processes are beyond the time-scale of molecular-dynamics
simulations. Further analysis of the excess volume remaining in the
nanocrystalline configurations revealed a transition from an extended
open pore network to closed pores at relative densities above 90\% for
all three metals.

In Ref.~\onlinecite{Derlet:01a} a model of nanocrystalline Ni based on
the Voronoi construction has been compared with an early version of
our model of nanocrystalline copper. This comparison showed that the
experimentally observed enhancement of the VDOS of nanocrystalline
materials at high and low energies is due to grain-boundary atoms,
whereas the influence of atoms at internal surfaces is negligible.  In
the present work we obtain the same result, using one and the same
model of nanocrystalline copper for all the calculations. Although we
find a small pressure to be present inside the grains of our model
systems, we do not observe a shift of the longitudinal modes in the
nanocrystalline material. A possible explanation for this difference
between the dense nanocrystalline material and the nanoparticle is the
loose coupling of the modes inside the grains via the grain
boundaries.

Our calculations of the specific heat of nanocrystalline Cu and a Cu
nanoparticle show that the low energy VDOS enhancements in the
nanoscale systems give rise to a pronounced peak in the excess
specific heat. This result is in qualitative and quantitative agreement
with other theoretical calculations\cite{Wolf:95a,Hu:01a} and at least
in qualitative agreement with experimental findings.\cite{Rupp:87a,
Tschoepe:93a,Loeffler:94a, Chen:95b,Bai:96a}

This work provides a deeper understanding of the vibrational
properties of nanoclusters and especially the differences between
these two nanoscale systems.  For future work it would be important to
investigate how the strong stress field present inside the
nanocrystalline model systems affects the vibrational properties as
well as some general study of the properties of crystalline systems
coupled by a disordered medium.  Finally, it would be interesting to
use some multi-scale method to relax our model systems beyond the time
scale of molecular-dynamics simulations.

%
%

\begin{acknowledgments}We would like to thank Prof. R. Birringer for the
provision of unpublished data of the specific heat in nanocrystalline
Pd.  This work has been supported by grants from the Canadian {Natural
Sciences and Engineering Research Council} (NSERC), Qu\'{e}bec's
\textit{Fonds qu\'{e}b\'{e}cois de la recherche sur la nature et les
technologies} (FQRNT, formerly FCAR) and Germany's \textit{Deutsche
Forschungsgmeinschaft} (DFG). Most of the calculations have been
performed on the facilities of the \textit{R\'{e}seau
qu\'{e}b\'{e}cois de calcul de haute performance} (RQCHP).
\end{acknowledgments}

%
%
\bibliography{nanovdos}

\begin{thebibliography}{54}
\expandafter\ifx\csname natexlab\endcsname\relax\def\natexlab#1{#1}\fi
\expandafter\ifx\csname bibnamefont\endcsname\relax
  \def\bibnamefont#1{#1}\fi
\expandafter\ifx\csname bibfnamefont\endcsname\relax
  \def\bibfnamefont#1{#1}\fi
\expandafter\ifx\csname citenamefont\endcsname\relax
  \def\citenamefont#1{#1}\fi
\expandafter\ifx\csname url\endcsname\relax
  \def\url#1{\texttt{#1}}\fi
\expandafter\ifx\csname urlprefix\endcsname\relax\def\urlprefix{URL }\fi
\providecommand{\bibinfo}[2]{#2}
\providecommand{\eprint}[2][]{\url{#2}}

\bibitem[{\citenamefont{Edelstein and Cammarata}(1996)}]{Edelstein:96a}
\bibinfo{editor}{\bibfnamefont{A.~S.} \bibnamefont{Edelstein}}
  \bibnamefont{and} \bibinfo{editor}{\bibfnamefont{R.~C.}
  \bibnamefont{Cammarata}}, eds., \emph{\bibinfo{title}{Nanomaterials:
  Synthesis, Properties and Applications}} (\bibinfo{publisher}{Institute of
  Physics Publishing}, \bibinfo{address}{Bristol}, \bibinfo{year}{1996}).

\bibitem[{\citenamefont{Siegel et~al.}(1988)\citenamefont{Siegel, Ramasamy,
  Hahn, Zongquan, Ting, and Gronsky}}]{Siegel:88a}
\bibinfo{author}{\bibfnamefont{R.~W.} \bibnamefont{Siegel}},
  \bibinfo{author}{\bibfnamefont{S.}~\bibnamefont{Ramasamy}},
  \bibinfo{author}{\bibfnamefont{H.}~\bibnamefont{Hahn}},
  \bibinfo{author}{\bibfnamefont{L.}~\bibnamefont{Zongquan}},
  \bibinfo{author}{\bibfnamefont{L.}~\bibnamefont{Ting}}, \bibnamefont{and}
  \bibinfo{author}{\bibfnamefont{R.}~\bibnamefont{Gronsky}},
  \bibinfo{journal}{J. Mater. Res.} \textbf{\bibinfo{volume}{3}},
  \bibinfo{pages}{1367} (\bibinfo{year}{1988}).

\bibitem[{\citenamefont{Kara and Rahman}(1998)}]{Kara:98a}
\bibinfo{author}{\bibfnamefont{A.}~\bibnamefont{Kara}} \bibnamefont{and}
  \bibinfo{author}{\bibfnamefont{T.~S.} \bibnamefont{Rahman}},
  \bibinfo{journal}{Phys. Rev. Lett.} \textbf{\bibinfo{volume}{81}},
  \bibinfo{pages}{1453} (\bibinfo{year}{1998}).

\bibitem[{\citenamefont{Hu et~al.}(2001)\citenamefont{Hu, Wang, Wu, Jiang, and
  Zi}}]{Hu:01a}
\bibinfo{author}{\bibfnamefont{X.}~\bibnamefont{Hu}},
  \bibinfo{author}{\bibfnamefont{G.}~\bibnamefont{Wang}},
  \bibinfo{author}{\bibfnamefont{W.}~\bibnamefont{Wu}},
  \bibinfo{author}{\bibfnamefont{P.}~\bibnamefont{Jiang}}, \bibnamefont{and}
  \bibinfo{author}{\bibfnamefont{J.}~\bibnamefont{Zi}}, \bibinfo{journal}{J.
  Phys.: Condens. Matter} \textbf{\bibinfo{volume}{13}}, \bibinfo{pages}{L835}
  (\bibinfo{year}{2001}).

\bibitem[{\citenamefont{Cleri and Rosato}(1993)}]{Cleri:93a}
\bibinfo{author}{\bibfnamefont{F.}~\bibnamefont{Cleri}} \bibnamefont{and}
  \bibinfo{author}{\bibfnamefont{V.}~\bibnamefont{Rosato}},
  \bibinfo{journal}{Phys. Rev. B} \textbf{\bibinfo{volume}{48}},
  \bibinfo{pages}{22} (\bibinfo{year}{1993}).

\bibitem[{\citenamefont{Phillpot
  et~al.}(1995{\natexlab{a}})\citenamefont{Phillpot, Wang, Wolf, and
  Gleiter}}]{Phillpot:95b}
\bibinfo{author}{\bibfnamefont{S.~R.} \bibnamefont{Phillpot}},
  \bibinfo{author}{\bibfnamefont{J.}~\bibnamefont{Wang}},
  \bibinfo{author}{\bibfnamefont{D.}~\bibnamefont{Wolf}}, \bibnamefont{and}
  \bibinfo{author}{\bibfnamefont{H.}~\bibnamefont{Gleiter}},
  \bibinfo{journal}{Mat. Sci. Eng. A} \textbf{\bibinfo{volume}{204}},
  \bibinfo{pages}{76} (\bibinfo{year}{1995}{\natexlab{a}}).

\bibitem[{\citenamefont{Phillpot
  et~al.}(1995{\natexlab{b}})\citenamefont{Phillpot, Wolf, and
  Gleiter}}]{Phillpot:95c}
\bibinfo{author}{\bibfnamefont{S.~R.} \bibnamefont{Phillpot}},
  \bibinfo{author}{\bibfnamefont{D.}~\bibnamefont{Wolf}}, \bibnamefont{and}
  \bibinfo{author}{\bibfnamefont{H.}~\bibnamefont{Gleiter}},
  \bibinfo{journal}{J. Appl. Phys.} \textbf{\bibinfo{volume}{78}},
  \bibinfo{pages}{847} (\bibinfo{year}{1995}{\natexlab{b}}).

\bibitem[{\citenamefont{Chen}(1995)}]{Chen:95a}
\bibinfo{author}{\bibfnamefont{D.}~\bibnamefont{Chen}}, \bibinfo{journal}{Comp.
  Mat. Sci.} \textbf{\bibinfo{volume}{3}}, \bibinfo{pages}{327}
  (\bibinfo{year}{1995}).

\bibitem[{\citenamefont{Schi{\o}tz et~al.}(1998)\citenamefont{Schi{\o}tz, {Di
  Tolla}, and Jacobsen}}]{Schiotz:98a}
\bibinfo{author}{\bibfnamefont{J.}~\bibnamefont{Schi{\o}tz}},
  \bibinfo{author}{\bibfnamefont{F.~D.} \bibnamefont{{Di Tolla}}},
  \bibnamefont{and} \bibinfo{author}{\bibfnamefont{K.~W.}
  \bibnamefont{Jacobsen}}, \bibinfo{journal}{Nature}
  \textbf{\bibinfo{volume}{391}}, \bibinfo{pages}{561} (\bibinfo{year}{1998}).

\bibitem[{\citenamefont{Schi{\o}tz et~al.}(1999)\citenamefont{Schi{\o}tz,
  Vegge, {Di Tolla}, and Jacobsen}}]{Schiotz:99a}
\bibinfo{author}{\bibfnamefont{J.}~\bibnamefont{Schi{\o}tz}},
  \bibinfo{author}{\bibfnamefont{T.}~\bibnamefont{Vegge}},
  \bibinfo{author}{\bibfnamefont{F.~D.} \bibnamefont{{Di Tolla}}},
  \bibnamefont{and} \bibinfo{author}{\bibfnamefont{K.~W.}
  \bibnamefont{Jacobsen}}, \bibinfo{journal}{Phys. Rev. B}
  \textbf{\bibinfo{volume}{60}}, \bibinfo{pages}{11971} (\bibinfo{year}{1999}).

\bibitem[{\citenamefont{{Van Swygenhoven} and Caro}(1997)}]{VanSwygenhoven:97a}
\bibinfo{author}{\bibfnamefont{H.}~\bibnamefont{{Van Swygenhoven}}}
  \bibnamefont{and} \bibinfo{author}{\bibfnamefont{A.}~\bibnamefont{Caro}},
  \bibinfo{journal}{Appl. Phys. Lett.} \textbf{\bibinfo{volume}{71}},
  \bibinfo{pages}{1652} (\bibinfo{year}{1997}).

\bibitem[{\citenamefont{{Van Swygenhoven} and Caro}(1998)}]{VanSwygenhoven:98a}
\bibinfo{author}{\bibfnamefont{H.}~\bibnamefont{{Van Swygenhoven}}}
  \bibnamefont{and} \bibinfo{author}{\bibfnamefont{A.}~\bibnamefont{Caro}},
  \bibinfo{journal}{Phys. Rev. B} \textbf{\bibinfo{volume}{58}},
  \bibinfo{pages}{11246} (\bibinfo{year}{1998}).

\bibitem[{\citenamefont{Nieman et~al.}(1991)\citenamefont{Nieman, Weertman, and
  Siegel}}]{Nieman:91a}
\bibinfo{author}{\bibfnamefont{G.~W.} \bibnamefont{Nieman}},
  \bibinfo{author}{\bibfnamefont{J.~R.} \bibnamefont{Weertman}},
  \bibnamefont{and} \bibinfo{author}{\bibfnamefont{R.~W.}
  \bibnamefont{Siegel}}, \bibinfo{journal}{J. Mater. Res.}
  \textbf{\bibinfo{volume}{6}}, \bibinfo{pages}{1012} (\bibinfo{year}{1991}).

\bibitem[{\citenamefont{Siegel}(1996)}]{Siegel:96a}
\bibinfo{author}{\bibfnamefont{R.~W.} \bibnamefont{Siegel}}, in
  \emph{\bibinfo{booktitle}{Nanomaterials, Synthesis, Properties and
  Applications}}, edited by \bibinfo{editor}{\bibfnamefont{A.~S.}
  \bibnamefont{Edelstein}} \bibnamefont{and}
  \bibinfo{editor}{\bibfnamefont{R.~C.} \bibnamefont{Camarata}}
  (\bibinfo{publisher}{Institute of Physics Publishing},
  \bibinfo{address}{Bristol}, \bibinfo{year}{1996}),
  chap.~\bibinfo{chapter}{9}.

\bibitem[{\citenamefont{Liu et~al.}(1994)\citenamefont{Liu, Adams, and
  Siegel}}]{Liu:94a}
\bibinfo{author}{\bibfnamefont{C.-L.} \bibnamefont{Liu}},
  \bibinfo{author}{\bibfnamefont{J.~B.} \bibnamefont{Adams}}, \bibnamefont{and}
  \bibinfo{author}{\bibfnamefont{R.~W.} \bibnamefont{Siegel}},
  \bibinfo{journal}{Nanostruct. Mater.} \textbf{\bibinfo{volume}{4}},
  \bibinfo{pages}{265} (\bibinfo{year}{1994}).

\bibitem[{\citenamefont{Zhu and Averback}(1995)}]{Zhu:95a}
\bibinfo{author}{\bibfnamefont{H.}~\bibnamefont{Zhu}} \bibnamefont{and}
  \bibinfo{author}{\bibfnamefont{R.~S.} \bibnamefont{Averback}},
  \bibinfo{journal}{Mater. Sci. Eng. A} \textbf{\bibinfo{volume}{204}},
  \bibinfo{pages}{96} (\bibinfo{year}{1995}).

\bibitem[{\citenamefont{Trampenau et~al.}(1995)\citenamefont{Trampenau,
  Bauszus, Petry, and Herr}}]{Trampenau:95a}
\bibinfo{author}{\bibfnamefont{J.}~\bibnamefont{Trampenau}},
  \bibinfo{author}{\bibfnamefont{K.}~\bibnamefont{Bauszus}},
  \bibinfo{author}{\bibfnamefont{W.}~\bibnamefont{Petry}}, \bibnamefont{and}
  \bibinfo{author}{\bibfnamefont{U.}~\bibnamefont{Herr}},
  \bibinfo{journal}{Nanostruct. Mater.} \textbf{\bibinfo{volume}{6}},
  \bibinfo{pages}{551} (\bibinfo{year}{1995}).

\bibitem[{\citenamefont{Fultz et~al.}(1996)\citenamefont{Fultz, Robertson,
  Stephens, Nagel, and Spooner}}]{Fultz:96a}
\bibinfo{author}{\bibfnamefont{B.}~\bibnamefont{Fultz}},
  \bibinfo{author}{\bibfnamefont{J.~L.} \bibnamefont{Robertson}},
  \bibinfo{author}{\bibfnamefont{T.~A.} \bibnamefont{Stephens}},
  \bibinfo{author}{\bibfnamefont{L.~J.} \bibnamefont{Nagel}}, \bibnamefont{and}
  \bibinfo{author}{\bibfnamefont{S.}~\bibnamefont{Spooner}},
  \bibinfo{journal}{J. Appl. Phys.} \textbf{\bibinfo{volume}{79}},
  \bibinfo{pages}{8318} (\bibinfo{year}{1996}).

\bibitem[{\citenamefont{Frase et~al.}(1998)\citenamefont{Frase, Fultz, and
  Robertson}}]{Frase:98a}
\bibinfo{author}{\bibfnamefont{H.}~\bibnamefont{Frase}},
  \bibinfo{author}{\bibfnamefont{B.}~\bibnamefont{Fultz}}, \bibnamefont{and}
  \bibinfo{author}{\bibfnamefont{J.~L.} \bibnamefont{Robertson}},
  \bibinfo{journal}{Phys. Rev. B} \textbf{\bibinfo{volume}{57}},
  \bibinfo{pages}{898} (\bibinfo{year}{1998}).

\bibitem[{\citenamefont{Bonetti et~al.}(2000)\citenamefont{Bonetti, Pasquini,
  Sampaolesi, Deriu, and Cicognani}}]{Bonetti:00a}
\bibinfo{author}{\bibfnamefont{E.}~\bibnamefont{Bonetti}},
  \bibinfo{author}{\bibfnamefont{L.}~\bibnamefont{Pasquini}},
  \bibinfo{author}{\bibfnamefont{E.}~\bibnamefont{Sampaolesi}},
  \bibinfo{author}{\bibfnamefont{A.}~\bibnamefont{Deriu}}, \bibnamefont{and}
  \bibinfo{author}{\bibfnamefont{G.}~\bibnamefont{Cicognani}},
  \bibinfo{journal}{J. Appl. Phys.} \textbf{\bibinfo{volume}{88}},
  \bibinfo{pages}{4571} (\bibinfo{year}{2000}).

\bibitem[{\citenamefont{Stuhr et~al.}(1998)\citenamefont{Stuhr, Wipf, Andersen,
  and Hahn}}]{Stuhr:98a}
\bibinfo{author}{\bibfnamefont{U.}~\bibnamefont{Stuhr}},
  \bibinfo{author}{\bibfnamefont{H.}~\bibnamefont{Wipf}},
  \bibinfo{author}{\bibfnamefont{K.~H.} \bibnamefont{Andersen}},
  \bibnamefont{and} \bibinfo{author}{\bibfnamefont{H.}~\bibnamefont{Hahn}},
  \bibinfo{journal}{Phys. Rev. Lett.} \textbf{\bibinfo{volume}{81}},
  \bibinfo{pages}{1449} (\bibinfo{year}{1998}).

\bibitem[{\citenamefont{Wolf et~al.}(1995)\citenamefont{Wolf, Wang, Phillpot,
  and Gleiter}}]{Wolf:95a}
\bibinfo{author}{\bibfnamefont{D.}~\bibnamefont{Wolf}},
  \bibinfo{author}{\bibfnamefont{J.}~\bibnamefont{Wang}},
  \bibinfo{author}{\bibfnamefont{S.~R.} \bibnamefont{Phillpot}},
  \bibnamefont{and} \bibinfo{author}{\bibfnamefont{H.}~\bibnamefont{Gleiter}},
  \bibinfo{journal}{Phys. Rev. Lett.} \textbf{\bibinfo{volume}{74}},
  \bibinfo{pages}{4686} (\bibinfo{year}{1995}).

\bibitem[{\citenamefont{Derlet et~al.}(2001)\citenamefont{Derlet, Meyer, Lewis,
  Stuhr, and {Van Swygenhoven}}}]{Derlet:01a}
\bibinfo{author}{\bibfnamefont{P.~M.} \bibnamefont{Derlet}},
  \bibinfo{author}{\bibfnamefont{R.}~\bibnamefont{Meyer}},
  \bibinfo{author}{\bibfnamefont{L.~J.} \bibnamefont{Lewis}},
  \bibinfo{author}{\bibfnamefont{U.}~\bibnamefont{Stuhr}}, \bibnamefont{and}
  \bibinfo{author}{\bibfnamefont{H.}~\bibnamefont{{Van Swygenhoven}}},
  \bibinfo{journal}{Phys. Rev. Lett.} \textbf{\bibinfo{volume}{87}},
  \bibinfo{pages}{205501} (\bibinfo{year}{2001}).

\bibitem[{\citenamefont{Meyer et~al.}(2002)\citenamefont{Meyer, Prakash, and
  Entel}}]{Meyer:02a}
\bibinfo{author}{\bibfnamefont{R.}~\bibnamefont{Meyer}},
  \bibinfo{author}{\bibfnamefont{S.}~\bibnamefont{Prakash}}, \bibnamefont{and}
  \bibinfo{author}{\bibfnamefont{P.}~\bibnamefont{Entel}},
  \bibinfo{journal}{Phase Trans.} \textbf{\bibinfo{volume}{75}},
  \bibinfo{pages}{51} (\bibinfo{year}{2002}).

\bibitem[{\citenamefont{Honeycutt and Andersen}(1987)}]{Honeycutt:87a}
\bibinfo{author}{\bibfnamefont{J.~D.} \bibnamefont{Honeycutt}}
  \bibnamefont{and} \bibinfo{author}{\bibfnamefont{H.~C.}
  \bibnamefont{Andersen}}, \bibinfo{journal}{J. Phys. Chem.}
  \textbf{\bibinfo{volume}{91}}, \bibinfo{pages}{4950} (\bibinfo{year}{1987}).

\bibitem[{\citenamefont{Swaminarayan et~al.}(1994)\citenamefont{Swaminarayan,
  Najafabadi, and Srolovitz}}]{Swaminarayan:94a}
\bibinfo{author}{\bibfnamefont{S.}~\bibnamefont{Swaminarayan}},
  \bibinfo{author}{\bibfnamefont{R.}~\bibnamefont{Najafabadi}},
  \bibnamefont{and} \bibinfo{author}{\bibfnamefont{D.~J.}
  \bibnamefont{Srolovitz}}, \bibinfo{journal}{Surf. Sci.}
  \textbf{\bibinfo{volume}{306}}, \bibinfo{pages}{367} (\bibinfo{year}{1994}).

\bibitem[{\citenamefont{L\'{o}pez et~al.}(1995)\citenamefont{L\'{o}pez, Marcos,
  and Alonso}}]{Lopez:95a}
\bibinfo{author}{\bibfnamefont{M.~J.} \bibnamefont{L\'{o}pez}},
  \bibinfo{author}{\bibfnamefont{P.~A.} \bibnamefont{Marcos}},
  \bibnamefont{and} \bibinfo{author}{\bibfnamefont{J.~A.}
  \bibnamefont{Alonso}}, \bibinfo{journal}{J. Chem. Phys.}
  \textbf{\bibinfo{volume}{104}}, \bibinfo{pages}{1056} (\bibinfo{year}{1995}).

\bibitem[{\citenamefont{Celino et~al.}(1996)\citenamefont{Celino, Cleri,
  D'Agostino, and Rosato}}]{Celino:96a}
\bibinfo{author}{\bibfnamefont{M.}~\bibnamefont{Celino}},
  \bibinfo{author}{\bibfnamefont{F.}~\bibnamefont{Cleri}},
  \bibinfo{author}{\bibfnamefont{G.}~\bibnamefont{D'Agostino}},
  \bibnamefont{and} \bibinfo{author}{\bibfnamefont{V.}~\bibnamefont{Rosato}},
  \bibinfo{journal}{Phys. Rev. Lett.} \textbf{\bibinfo{volume}{77}},
  \bibinfo{pages}{2495} (\bibinfo{year}{1996}).

\bibitem[{\citenamefont{Palacios et~al.}(1999)\citenamefont{Palacios,
  {I\~{n}iguez}, L\'{o}pez, and Alonso}}]{Palacios:99a}
\bibinfo{author}{\bibfnamefont{F.~J.} \bibnamefont{Palacios}},
  \bibinfo{author}{\bibfnamefont{M.~P.} \bibnamefont{{I\~{n}iguez}}},
  \bibinfo{author}{\bibfnamefont{M.~J.} \bibnamefont{L\'{o}pez}},
  \bibnamefont{and} \bibinfo{author}{\bibfnamefont{J.~A.}
  \bibnamefont{Alonso}}, \bibinfo{journal}{Phys. Rev. B}
  \textbf{\bibinfo{volume}{60}}, \bibinfo{pages}{2908} (\bibinfo{year}{1999}).

\bibitem[{\citenamefont{Sun et~al.}(2001)\citenamefont{Sun, Ren, Luo, and
  Wang}}]{Sun:01a}
\bibinfo{author}{\bibfnamefont{H.~Q.} \bibnamefont{Sun}},
  \bibinfo{author}{\bibfnamefont{Y.}~\bibnamefont{Ren}},
  \bibinfo{author}{\bibfnamefont{Y.~H.} \bibnamefont{Luo}}, \bibnamefont{and}
  \bibinfo{author}{\bibfnamefont{G.~H.} \bibnamefont{Wang}},
  \bibinfo{journal}{Physica B} \textbf{\bibinfo{volume}{293}},
  \bibinfo{pages}{260} (\bibinfo{year}{2001}).

\bibitem[{\citenamefont{Darby et~al.}(2002)\citenamefont{Darby, Mortimer-Jones,
  Johnston, and Roberts}}]{Darby:02a}
\bibinfo{author}{\bibfnamefont{S.}~\bibnamefont{Darby}},
  \bibinfo{author}{\bibfnamefont{T.~V.} \bibnamefont{Mortimer-Jones}},
  \bibinfo{author}{\bibfnamefont{R.~L.} \bibnamefont{Johnston}},
  \bibnamefont{and} \bibinfo{author}{\bibfnamefont{C.}~\bibnamefont{Roberts}},
  \bibinfo{journal}{J. Chem. Phys.} \textbf{\bibinfo{volume}{116}},
  \bibinfo{pages}{1536} (\bibinfo{year}{2002}).

\bibitem[{\citenamefont{Michaelian et~al.}(2002)\citenamefont{Michaelian,
  Beltran, and Garzon}}]{Michaelian:02a}
\bibinfo{author}{\bibfnamefont{K.}~\bibnamefont{Michaelian}},
  \bibinfo{author}{\bibfnamefont{M.~R.} \bibnamefont{Beltran}},
  \bibnamefont{and} \bibinfo{author}{\bibfnamefont{I.~L.}
  \bibnamefont{Garzon}}, \bibinfo{journal}{Phys. Rev. B}
  \textbf{\bibinfo{volume}{65}}, \bibinfo{pages}{041403}
  (\bibinfo{year}{2002}).

\bibitem[{\citenamefont{Rexer et~al.}(2002)\citenamefont{Rexer, Jellinek,
  Krissinel, Parks, and Riley}}]{Rexer:02a}
\bibinfo{author}{\bibfnamefont{E.~F.} \bibnamefont{Rexer}},
  \bibinfo{author}{\bibfnamefont{J.}~\bibnamefont{Jellinek}},
  \bibinfo{author}{\bibfnamefont{E.~B.} \bibnamefont{Krissinel}},
  \bibinfo{author}{\bibfnamefont{E.~K.} \bibnamefont{Parks}}, \bibnamefont{and}
  \bibinfo{author}{\bibfnamefont{S.~J.} \bibnamefont{Riley}},
  \bibinfo{journal}{J. Chem. Phys.} \textbf{\bibinfo{volume}{117}},
  \bibinfo{pages}{82} (\bibinfo{year}{2002}).

\bibitem[{\citenamefont{Aguilera-Granja
  et~al.}(2002)\citenamefont{Aguilera-Granja, Rodr\'{i}guez-L\'{o}pez,
  Michaelian, Berlanga-Ram\'{i}rez, and Vega}}]{AguileraGranja:02a}
\bibinfo{author}{\bibfnamefont{F.}~\bibnamefont{Aguilera-Granja}},
  \bibinfo{author}{\bibfnamefont{J.~L.} \bibnamefont{Rodr\'{i}guez-L\'{o}pez}},
  \bibinfo{author}{\bibfnamefont{K.}~\bibnamefont{Michaelian}},
  \bibinfo{author}{\bibfnamefont{E.~O.} \bibnamefont{Berlanga-Ram\'{i}rez}},
  \bibnamefont{and} \bibinfo{author}{\bibfnamefont{A.}~\bibnamefont{Vega}},
  \bibinfo{journal}{Phys. Rev. B} \textbf{\bibinfo{volume}{66}},
  \bibinfo{pages}{224410} (\bibinfo{year}{2002}).

\bibitem[{\citenamefont{Celino et~al.}(1995{\natexlab{a}})\citenamefont{Celino,
  D'Agostino, and Rosato}}]{Celino:95a}
\bibinfo{author}{\bibfnamefont{M.}~\bibnamefont{Celino}},
  \bibinfo{author}{\bibfnamefont{G.}~\bibnamefont{D'Agostino}},
  \bibnamefont{and} \bibinfo{author}{\bibfnamefont{V.}~\bibnamefont{Rosato}},
  \bibinfo{journal}{Mater. Sci. Eng. A} \textbf{\bibinfo{volume}{201}},
  \bibinfo{pages}{101} (\bibinfo{year}{1995}{\natexlab{a}}).

\bibitem[{\citenamefont{Celino et~al.}(1995{\natexlab{b}})\citenamefont{Celino,
  D'Agostino, and Rosato}}]{Celino:95b}
\bibinfo{author}{\bibfnamefont{M.}~\bibnamefont{Celino}},
  \bibinfo{author}{\bibfnamefont{G.}~\bibnamefont{D'Agostino}},
  \bibnamefont{and} \bibinfo{author}{\bibfnamefont{V.}~\bibnamefont{Rosato}},
  \bibinfo{journal}{Nanostruct. Mater.} \textbf{\bibinfo{volume}{6}},
  \bibinfo{pages}{751} (\bibinfo{year}{1995}{\natexlab{b}}).

\bibitem[{\citenamefont{Derlet and {Van Swygenhoven}}(2002)}]{Derlet:02a}
\bibinfo{author}{\bibfnamefont{P.~M.} \bibnamefont{Derlet}} \bibnamefont{and}
  \bibinfo{author}{\bibfnamefont{H.}~\bibnamefont{{Van Swygenhoven}}},
  \bibinfo{journal}{Phil. Mag. A} \textbf{\bibinfo{volume}{82}},
  \bibinfo{pages}{1} (\bibinfo{year}{2002}).

\bibitem[{\citenamefont{Samaras et~al.}(2002)\citenamefont{Samaras, Derlet,
  {Van Swygenhoven}, and Victoria}}]{Samaras:02a}
\bibinfo{author}{\bibfnamefont{M.}~\bibnamefont{Samaras}},
  \bibinfo{author}{\bibfnamefont{P.~M.} \bibnamefont{Derlet}},
  \bibinfo{author}{\bibfnamefont{H.}~\bibnamefont{{Van Swygenhoven}}},
  \bibnamefont{and} \bibinfo{author}{\bibfnamefont{M.}~\bibnamefont{Victoria}},
  \bibinfo{journal}{Phys. Rev. Lett.} \textbf{\bibinfo{volume}{88}},
  \bibinfo{pages}{125505} (\bibinfo{year}{2002}).

\bibitem[{\citenamefont{Allen and Tildesley}(1991)}]{Allen:91a}
\bibinfo{author}{\bibfnamefont{M.~P.} \bibnamefont{Allen}} \bibnamefont{and}
  \bibinfo{author}{\bibfnamefont{D.~J.} \bibnamefont{Tildesley}},
  \emph{\bibinfo{title}{Computer Simulations of Liquids}}
  (\bibinfo{publisher}{Clarendon}, \bibinfo{address}{Oxford},
  \bibinfo{year}{1991}).

\bibitem[{\citenamefont{Weeks et~al.}(1971)\citenamefont{Weeks, Chandler, and
  Andersen}}]{Weeks:71a}
\bibinfo{author}{\bibfnamefont{J.}~\bibnamefont{Weeks}},
  \bibinfo{author}{\bibfnamefont{D.}~\bibnamefont{Chandler}}, \bibnamefont{and}
  \bibinfo{author}{\bibfnamefont{H.}~\bibnamefont{Andersen}},
  \bibinfo{journal}{J. Chem. Phys.} \textbf{\bibinfo{volume}{54}},
  \bibinfo{pages}{5237} (\bibinfo{year}{1971}).

\bibitem[{\citenamefont{Parrinello and Rahman}(1980)}]{Parrinello:80a}
\bibinfo{author}{\bibfnamefont{M.}~\bibnamefont{Parrinello}} \bibnamefont{and}
  \bibinfo{author}{\bibfnamefont{A.}~\bibnamefont{Rahman}},
  \bibinfo{journal}{Phys. Rev. Lett.} \textbf{\bibinfo{volume}{45}},
  \bibinfo{pages}{1196} (\bibinfo{year}{1980}).

\bibitem[{\citenamefont{Hoover}(1985)}]{Hoover:85a}
\bibinfo{author}{\bibfnamefont{W.~G.} \bibnamefont{Hoover}},
  \bibinfo{journal}{Phys. Rev. A} \textbf{\bibinfo{volume}{31}},
  \bibinfo{pages}{1695} (\bibinfo{year}{1985}).

\bibitem[{\citenamefont{Campbell et~al.}(1999)\citenamefont{Campbell, Kalia,
  Nakano, Shimojo, Tsuruta, Vashishta, and Ogata}}]{Campbell:99a}
\bibinfo{author}{\bibfnamefont{T.}~\bibnamefont{Campbell}},
  \bibinfo{author}{\bibfnamefont{R.~K.} \bibnamefont{Kalia}},
  \bibinfo{author}{\bibfnamefont{A.}~\bibnamefont{Nakano}},
  \bibinfo{author}{\bibfnamefont{F.}~\bibnamefont{Shimojo}},
  \bibinfo{author}{\bibfnamefont{K.}~\bibnamefont{Tsuruta}},
  \bibinfo{author}{\bibfnamefont{P.}~\bibnamefont{Vashishta}},
  \bibnamefont{and} \bibinfo{author}{\bibfnamefont{S.}~\bibnamefont{Ogata}},
  \bibinfo{journal}{Phys. Rev. Lett.} \textbf{\bibinfo{volume}{82}},
  \bibinfo{pages}{4018} (\bibinfo{year}{1999}).

\bibitem[{\citenamefont{Wasserman and Vermaak}(1970)}]{Wasserman:70a}
\bibinfo{author}{\bibfnamefont{H.~J.} \bibnamefont{Wasserman}}
  \bibnamefont{and} \bibinfo{author}{\bibfnamefont{J.~S.}
  \bibnamefont{Vermaak}}, \bibinfo{journal}{Surf. Sci}
  \textbf{\bibinfo{volume}{22}}, \bibinfo{pages}{164} (\bibinfo{year}{1970}).

\bibitem[{\citenamefont{Mays et~al.}(1968)\citenamefont{Mays, Vermaak, and
  Kuhlmann-Wilsdorf}}]{Mays:68b}
\bibinfo{author}{\bibfnamefont{C.~W.} \bibnamefont{Mays}},
  \bibinfo{author}{\bibfnamefont{J.~S.} \bibnamefont{Vermaak}},
  \bibnamefont{and}
  \bibinfo{author}{\bibfnamefont{D.}~\bibnamefont{Kuhlmann-Wilsdorf}},
  \bibinfo{journal}{Surf. Sci} \textbf{\bibinfo{volume}{12}},
  \bibinfo{pages}{134} (\bibinfo{year}{1968}).

\bibitem[{\citenamefont{Solliard and Flueli}(1985)}]{Solliard:85a}
\bibinfo{author}{\bibfnamefont{C.}~\bibnamefont{Solliard}} \bibnamefont{and}
  \bibinfo{author}{\bibfnamefont{M.}~\bibnamefont{Flueli}},
  \bibinfo{journal}{Surf. Sci.} \textbf{\bibinfo{volume}{156}},
  \bibinfo{pages}{487} (\bibinfo{year}{1985}).

\bibitem[{\citenamefont{Wasserman and Vermaak}(1972)}]{Wasserman:72a}
\bibinfo{author}{\bibfnamefont{H.~J.} \bibnamefont{Wasserman}}
  \bibnamefont{and} \bibinfo{author}{\bibfnamefont{J.~S.}
  \bibnamefont{Vermaak}}, \bibinfo{journal}{Surf. Sci}
  \textbf{\bibinfo{volume}{32}}, \bibinfo{pages}{168} (\bibinfo{year}{1972}).

\bibitem[{\citenamefont{Black}(1990)}]{Black:90a}
\bibinfo{author}{\bibfnamefont{J.~E.} \bibnamefont{Black}}, in
  \emph{\bibinfo{booktitle}{Dynamical Properties of Solids}}, edited by
  \bibinfo{editor}{\bibfnamefont{G.~K.} \bibnamefont{Horton}} \bibnamefont{and}
  \bibinfo{editor}{\bibfnamefont{A.~A.} \bibnamefont{Maradudin}}
  (\bibinfo{publisher}{North Holland}, \bibinfo{address}{Amsterdam},
  \bibinfo{year}{1990}), vol.~\bibinfo{volume}{6}.

\bibitem[{\citenamefont{Meyer and Lewis}(2002)}]{Meyer:02b}
\bibinfo{author}{\bibfnamefont{R.}~\bibnamefont{Meyer}} \bibnamefont{and}
  \bibinfo{author}{\bibfnamefont{L.~J.} \bibnamefont{Lewis}},
  \bibinfo{journal}{Phys. Rev. B} \textbf{\bibinfo{volume}{66}},
  \bibinfo{pages}{052106} (\bibinfo{year}{2002}).

\bibitem[{\citenamefont{Rupp and Birringer}(1987)}]{Rupp:87a}
\bibinfo{author}{\bibfnamefont{J.}~\bibnamefont{Rupp}} \bibnamefont{and}
  \bibinfo{author}{\bibfnamefont{R.}~\bibnamefont{Birringer}},
  \bibinfo{journal}{Phys. Rev. B} \textbf{\bibinfo{volume}{36}},
  \bibinfo{pages}{7888} (\bibinfo{year}{1987}).

\bibitem[{\citenamefont{Tsch\"{o}pe and Birringer}(1993)}]{Tschoepe:93a}
\bibinfo{author}{\bibfnamefont{A.}~\bibnamefont{Tsch\"{o}pe}} \bibnamefont{and}
  \bibinfo{author}{\bibfnamefont{R.}~\bibnamefont{Birringer}},
  \bibinfo{journal}{Phil. Mag. B} \textbf{\bibinfo{volume}{68}},
  \bibinfo{pages}{223} (\bibinfo{year}{1993}).

\bibitem[{\citenamefont{L\"{o}ffler}(1994)}]{Loeffler:94a}
\bibinfo{author}{\bibfnamefont{J.}~\bibnamefont{L\"{o}ffler}},
  \emph{\bibinfo{title}{Diplomathesis}} (\bibinfo{publisher}{Universit\"{a}t
  des Saarlandes, Germany}, \bibinfo{year}{1994}).

\bibitem[{\citenamefont{Chen et~al.}(1995)\citenamefont{Chen, Yao, Hsiao, Jen,
  Lin, Lin, and Tung}}]{Chen:95b}
\bibinfo{author}{\bibfnamefont{Y.~Y.} \bibnamefont{Chen}},
  \bibinfo{author}{\bibfnamefont{Y.~D.} \bibnamefont{Yao}},
  \bibinfo{author}{\bibfnamefont{S.~S.} \bibnamefont{Hsiao}},
  \bibinfo{author}{\bibfnamefont{S.~U.} \bibnamefont{Jen}},
  \bibinfo{author}{\bibfnamefont{B.~T.} \bibnamefont{Lin}},
  \bibinfo{author}{\bibfnamefont{H.~M.} \bibnamefont{Lin}}, \bibnamefont{and}
  \bibinfo{author}{\bibfnamefont{C.~Y.} \bibnamefont{Tung}},
  \bibinfo{journal}{Phys. Rev. B} \textbf{\bibinfo{volume}{52}},
  \bibinfo{pages}{9364} (\bibinfo{year}{1995}).

\bibitem[{\citenamefont{Bai et~al.}(1996)\citenamefont{Bai, Luo, and
  Jin}}]{Bai:96a}
\bibinfo{author}{\bibfnamefont{H.~Y.} \bibnamefont{Bai}},
  \bibinfo{author}{\bibfnamefont{J.~L.} \bibnamefont{Luo}}, \bibnamefont{and}
  \bibinfo{author}{\bibfnamefont{D.}~\bibnamefont{Jin}}, \bibinfo{journal}{J.
  Appl. Phys.} \textbf{\bibinfo{volume}{79}}, \bibinfo{pages}{361}
  (\bibinfo{year}{1996}).

\end{thebibliography}
\end{document}